 \documentclass[twocolumn,superscriptaddress]{revtex4-1}
\usepackage[aps]{optional}

\usepackage{bm}
\usepackage{graphicx}
\usepackage{amsmath}

\usepackage[utf8]{inputenc}
\usepackage{hyperref}

\begin{document}
\title{Ultrastrong coupling between a microwave resonator and antiferromagnetic resonances of rare earth ion spins}

\opt{nat}{
\author{Jono Everts$^{1}$, Gavin King$^{1}$, Nicholas Lambert$^{1}$, Sascha Kocsis$^{2}$, Sven Rogge$^{2}$, Jevon J. Longdell$^{1}$}
\maketitle
\begin{affiliations}
\item The Dodd-Walls Centre for Photonic and Quantum Technologies, Department of Physics, University of Otago,
Dunedin, New Zealand
\item Centre for Quantum Computation and Communication Technology, The University of New South Wales, Sydney, New South Wales 2052, Australia
\end{affiliations}
}

\opt{aps}{
  \author{Jonathan Everts}
  \affiliation{The Dodd-Walls Centre for Photonic and Quantum Technologies, Department of Physics, University of Otago,
    Dunedin, New Zealand}
  \author{Gavin G. G. King}
  \affiliation{The Dodd-Walls Centre for Photonic and Quantum Technologies, Department of Physics, University of Otago,
    Dunedin, New Zealand}
  \author{Nicholas Lambert}
  \affiliation{The Dodd-Walls Centre for Photonic and Quantum Technologies, Department of Physics, University of Otago,
    Dunedin, New Zealand}
  \author{Sacha Kocsis}
  \affiliation{Centre for Quantum Computation and Communication Technology, The University of New South Wales, Sydney, New South Wales 2052, Australia}
   \author{Sven Rogge}
  \affiliation{Centre for Quantum Computation and Communication Technology, The University of New South Wales, Sydney, New South Wales 2052, Australia}
  \author{Jevon J. Longell}
  \affiliation{The Dodd-Walls Centre for Photonic and Quantum Technologies, Department of Physics, University of Otago,
    Dunedin, New Zealand}
}

\maketitle

\textbf{Quantum magnonics is a new and active research field, leveraging the strong collective coupling between microwaves and magnetically ordered spin systems \cite{lachance-quirion_hybrid_2019}. To date work in quantum magnonics has focused on transition metals and almost entirely on ferromagnetic resonances in yttrium iron garnet (YIG) \cite{tabuchi_hybridizing_2014, Goryachev2014}. Antiferromagnetic systems have gained interest as they produce no stray field, and are therefore robust to magnetic perturbations and have narrow, shape independent resonant linewidths \cite{Yuan_2017}. Here we show the first experimental evidence of ultrastrong-coupling between a microwave cavity and collective antiferromagnetic resonances (magnons) in a rare earth crystal. The combination of the unique optical and spin\cite{thiel_rare-earth-doped_2011,ahlefeldt_ultranarrow_2016} properties of the rare earths and collective antiferromagnetic order paves the way for novel quantum magnonic applications.}

 Superconducting qubits have provided a new set of capabilities in quantum computing, control and measurement. In turn, this has generated much interest in ``hybrid quantum systems'' \cite{kurizki_quantum_2015}. In such approaches the superconducting qubits capabilities are enhanced by coupling them to another type of system. Amongst the challenges that the hybrid approach tries to address is the long term storage of quantum information by using spins, as well as long distance, room temperature,  quantum communication by  microwave to optical conversion \cite{lambert_coherent_2019}.

Rare earth ions offer exciting possibilities for quantum information because of the long coherence times available for both their optical \cite{bottger_spectroscopy_2006} and optically addressed spin transitions \cite{zhong_optically_2015}. Quantum memories based on ensembles of rare earth ions have demonstrated large bandwidths \cite{clausen_quantum_2011}, high efficiencies and very long storage times \cite{zhong_optically_2015}. 

Rare earth ions in solids are also being investigated for microwave to optical conversion \cite{obrien_interfacing_2014,williamson_magneto-optic_2014,fernandez-gonzalvo_cavity-enhanced_2019, welinski_electron_2019, everts_microwave_2019}. When using rare earth doped samples for these applications, there is a tradeoff when it comes to the concentration of the dopant. Higher dopant concentrations cause a desirable increase in the collective coupling strength to electromagnetic waves but an undesirable increase in both homogeneous and inhomogeneous linewidths. For optical transitions in low concentration rare earth dopants, this increase is often due to Stark shifts or an increase in crystal strain from dopants degrading the crystal quality \cite{sellars_investigation_2004}. For electron spin transitions it is often due to magnetic dipole-dipole interactions \cite{biasi_esr_1983}.

A way out of this concentration-linewidth trade off is to instead use fully concentrated rare earth crystals where the rare earth is part of the host crystal. In the case of optical transitions it has been shown that very narrow inhomogeneous linewidths (25\,MHz)\cite{ahlefeldt_ultranarrow_2016} can be achieved in fully concentrated samples, linewidths comparable with the narrowest seen in doped samples \cite{macfarlane_optical_1998,thiel_rare-earth-doped_2011}.

Here we investigate the coupling between antiferromagnetic magnon modes in the fully concentrated rare earth crystal gadolinium vanadate (GdVO$_4$) and a microwave cavity. We show that not only can ultrastrong-coupling be achieved but that the linewidth of the magnon resonances are narrow, an important property for their use in hybrid quantum systems.

The magnetic and thermal properties of fully concentrated rare earth phosphates and vanadates have been investigated before \cite{Bleaney2000, Bowden1998}, and collective resonances were observed in gadolinium aluminate (GdAlO$_4$) and GdVO$_4$. These were observed however, at much higher temperatures and showed much larger linewidths than we observe here. In general antiferromagnetic resonance in rare earth systems is a largely unexplored area.


GdVO$_4$  has a tetragonal crystal lattice with space group $D^{19}_{4h}(I4/amd)$.  There are four Gd$^{3+}$ ions per unit cell with site symmetry $D_{2d}$. At room temperature GdVO$_4$ is paramagnetic but upon cooling at zero-magnetic field to $T \leq$~2.495\,K, it orders antiferromagnetically as a simple two sublattice system parallel to the crystallographic $c$-axis \cite{Cashion.1970}. The low ordering temperature is a characteristic of rare earth spins as the exchange interaction between the highly shielded $4f$ electrons is weak, and often the magnetic dipole-dipole interaction dominates \cite{lagendijk_caloric_1972}. The Gd$^{3+}$ ions are in an $^8$S$_{7/2}$ ground state, a predominantly $L = 0$ state; crystal field effects are therefore small and the g-factor is approximately isotropic with a value of 2. GdVO$_4$ is iso-structural with the ubiquitous laser host material yttrium vanadate and is a good laser host in its own right. The sample we used was commerically grown and available off the shelf.


The antiferromagnetic resonance \cite{Kittel.1951} seen in our system can be understood by modelling each sublattice by a single \mbox{spin-$1/2$} and using the following interaction Hamiltonian

\begin{equation}
  H = AS_1^zS_2^z - \mu_Bg \bm B\cdot (\bm S_1+\bm S_2).
  \label{eq:Hamiltonian1}
\end{equation}

The first term describes the (predominantly) magnetic dipole-dipole interaction that causes antiferromagnetic ordering and the second term describes the interaction of the spins with an applied magnetic field.

We take $S_1^z$ to be our upward pointing spin ($S_z^1\approx \frac{1}{2}$, $S_z^2\approx -\frac{1}{2}$), and apply a static magnetic field of strength $B_0$ along the $z$-axis. Assuming only small excitations of the spins we make the Holstein-Primakoff transformations \cite{Holstein1940}

\begin{align}
S_1^z &= \frac{1}{2} -\hat{a}^{\dagger}\hat{a} \\
S_2^z &= -\frac{1}{2} + \hat{b}^{\dagger}\hat{b} \\
S_1^+ &\approx  \hat{a} \\
S_2^- &\approx  \hat{b},
\end{align}
where $\hat{a}$ and $\hat{b}$ are bosonic annihilation operators for a spin excitation (spin flip) on sublattice 1 and 2 respectively. Substituting into Eq.\ref{eq:Hamiltonian1} and keeping terms up to second-order in creation/annihilation operators leads to the Hamiltonian

\begin{equation}
  H = \hbar \omega_0(\hat{a}^\dagger \hat{a} + \hat{b}^\dagger \hat{b}) + g\mu_BB_0(\hat{a}^\dagger \hat{a} - \hat{b}^\dagger \hat{b})\\, \label{eq:Hamiltonian2}
\end{equation}
where $\omega_0 = A/2\hbar$. Thus in our simple model there are two antiferromagnetic resonance modes each involving Larmor precession of just one sublattice. At zero magnetic field, the two modes are degenerate at a frequency that is directly related to the strength of the ordering interaction. The frequency of mode $\hat{a}$, which sees the applied magnetic field in addition to the internal field, increases linearly with applied magnetic field, where as the frequency of mode $\hat{b}$, which sees the applied field opposing the internal field, decreases linearly with applied magnetic field.

\begin{figure}
  \centering
  \includegraphics[width=1.0\columnwidth]{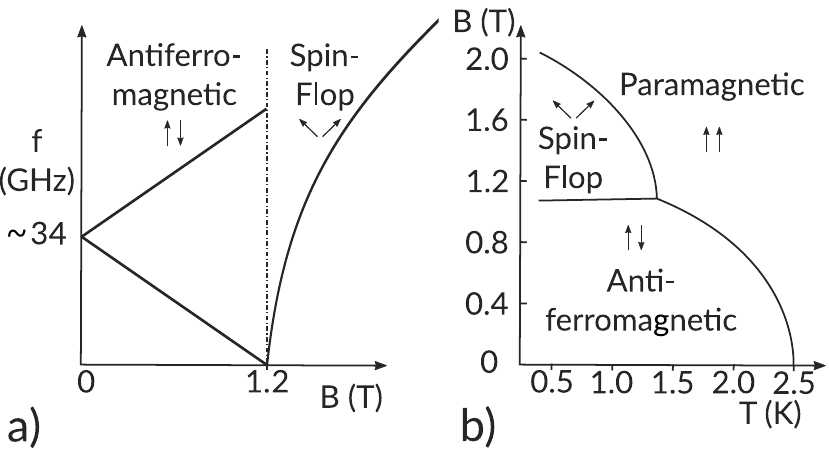}
  \caption{\label{fig:spinfrequencies}  (a) The antiferromagnetic resonance frequencies of GdVO$_4$ as a function of magnetic field. Beyond the spin-flop transition Eq.\ref{eq:Hamiltonian2} no longer holds, spin resonance theory in the spin-flop state is required instead \cite{Yungli_1964}.
  \label{fig:magneticphase} (b) The magnetic phase diagram of GdVO$_4$ as a function of applied magnetic field strength and temperature \cite{Mangum.1972}.
  }
\end{figure}

Antiferromagnetic resonance in GdVO$_4$ has been seen in electron spin resonance experiments, with a predicted low temperature zero-field frequency of $\sim34$\,GHz \cite{Abraham.1992}. Using a g-factor of 2 the antiferromagnetic resonance frequencies predicted by Eq.~\ref{eq:Hamiltonian2} are plotted in Fig.~\ref{fig:spinfrequencies}(a). At $\sim1.2$ T the lower antiferromagnetic resonance branch intercepts the zero frequency axis. At this point a transition occurs between the antiferromagnetic phase and the spin-flop phase. In the spin flop phase, the energy penalty associated with having one of the sublattices anti-aligned with the external field is relieved by the spins switching from the crystal $c$-axis to the crystal $a$-axis, and tilting a small angle $\theta$ towards the external field. As the external field is increased, $\theta$ increases, eventually returning the crystal to the paramagnetic phase. The magnetic phase in GdVO$_4$ is shown as a function of temperature and magnetic field in Fig.~\ref{fig:magneticphase}(b).

\begin{figure}
  \centering
\includegraphics[width=1.0\columnwidth]{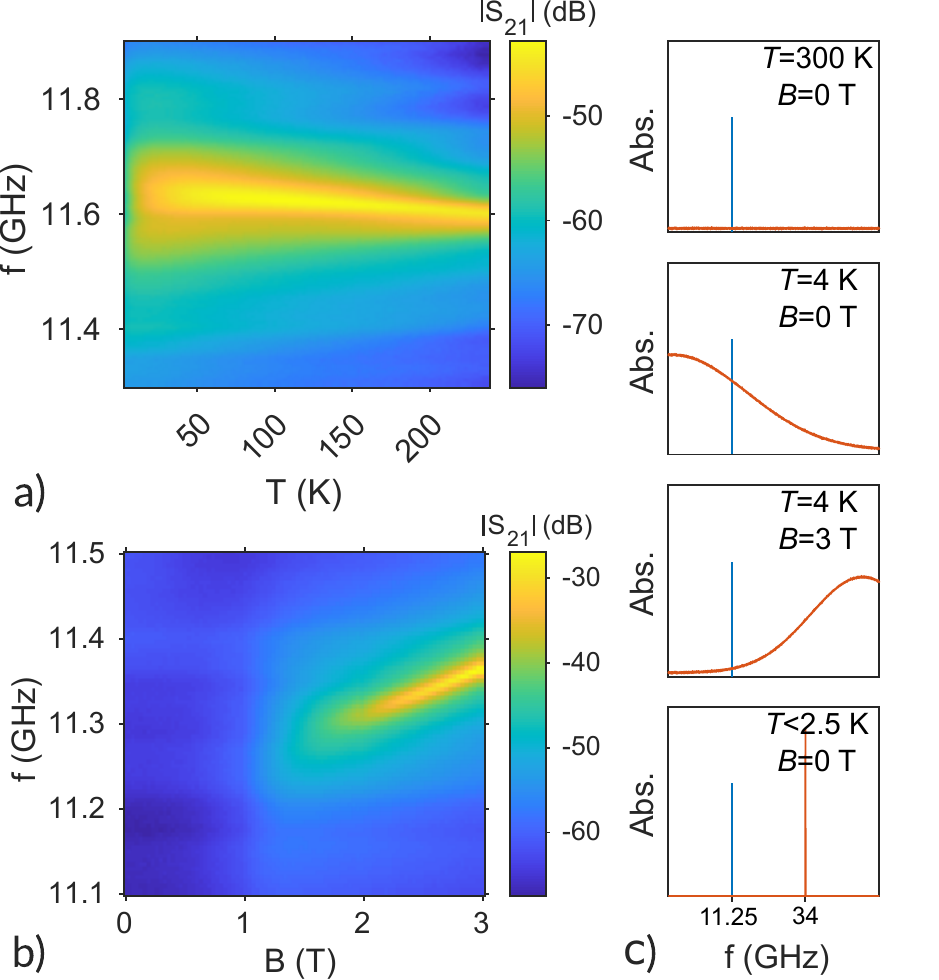}
    \caption{\label{fig:coolingto4K}(a) The cavity transmission as cooled from room temperature to 4\,K.
    \label{fig:4Kfieldsweep} (b) Cavity transmission as a function of applied magnetic field with a fixed temperature of $T =$ 4\,K. 
    \label{fig:cavitymodediagrams} (c) A sketch showing the relative locations of the cavity (blue) and  spin (orange) resonances at different fields and temperatures.}
\end{figure}

\begin{figure}
    \centering
    \includegraphics[width=\columnwidth]{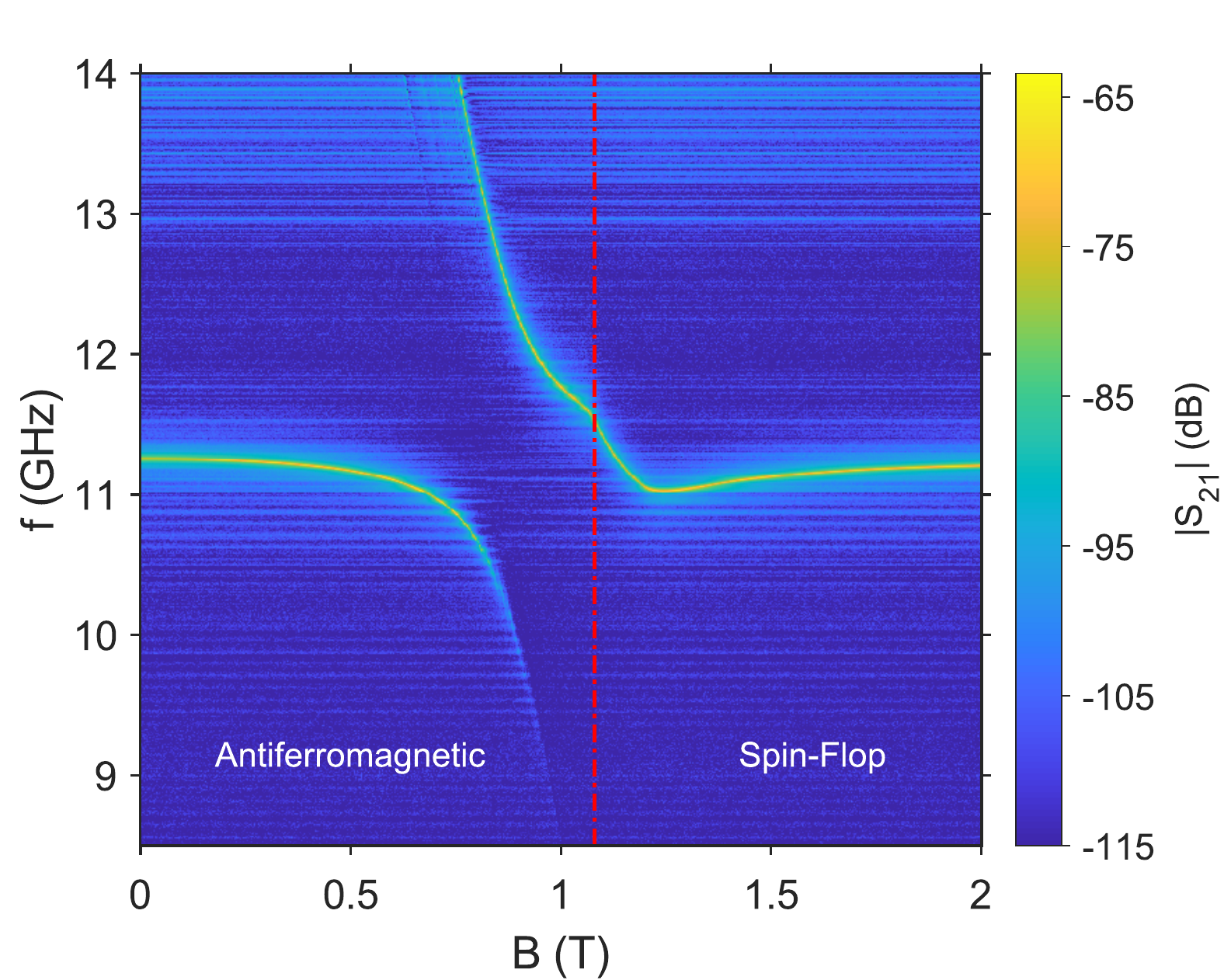}
    \caption{
  \label{fig:overallavoidedcrossing} $T = 25$\,mK, transmission through the cavity as a function of magnetic field and frequency. The red line indicates the onset of the spin-flop phase.}
\end{figure}

The experimental setup is shown in Fig.~\ref{fig:setup} the GdVO$_4$ sample was placed inside a loop-gap resonator and cooled within a dilution refrigerator. Figure~\ref{fig:coolingto4K}(a) shows the transmission through the cavity as the cavity is cooled from room temperature to 4\,K, with zero applied magnetic field. With decreasing temperature the resonance frequency of the cavity is seen to increase and broaden until the resonance completely disappears at $T \le 35$\,K. This occurs due to the interaction between the cavity and the spins. At 300\,K, there is a \emph{very} broad paramagnetic resonance centered on 0\,Hz. Because it is so broad there is very little absorbtion at the cavity frequency. As the temperature decreases, spin lattice relaxation decreases and the spins resonance narrows, leading to more loss at the cavity frequency. Once the temperature reaches 4\,K the attenuation has increased to the point that the cavity resonance disappears. 

Keeping the temperature fixed at 4\,K an external field is applied (along the crystals $c$-axis) to shift the paramagnetic resonance of the ions away from the cavity. In Fig.~\ref{fig:4Kfieldsweep}(b) we see that as the magnetic field increases and the resonance of the ions is moved, the source of loss to the cavity is reduced allowing the cavity resonance to reappear.

The cavity resonance also reappears when the sample is cooled past the transition temperature $T \le$~2.495\,K. Below the transition temperature the spins become locked together in a long-range order giving a narrow resonant peak at the zero field antiferromagnetic resonance frequency, $\sim34$\,GHz. The cavity is well detuned from this peak and hence resonates as if it were empty.

To investigate the coupling between the antiferromagnetic resonant mode and the microwave cavity, the cavity transmission is measured as the lower antiferromagnetic resonant branch (shown in Fig.~\ref{fig:spinfrequencies}) is pulled through the cavity resonance via sweeping the applied magnetic field. With the temperature fixed at 25\,mK  Fig.~\ref{fig:overallavoidedcrossing}(a) shows the cavity transmission as a function of applied magnetic field. As the lower antiferromagnetic resonant branch passes through the cavity resonance a clear avoided crossing is seen centered at $\approx0.9$\,T.
The shape is not symmetric, there is a significant bump in the dressed state frequency at $\approx1.1$\,T due to the onset of the spin-flop transition. A detailed investigation of this phenomenon will be left to future work.

We fit to our data (for $B<1.1$\,T)  the eigenvalues of a simple coupling Hamiltonian 

\begin{equation}
    H = \omega_c \hat{c}^{\dagger}\hat{c} + \omega_m\hat{m}^{\dagger}\hat{m} + G\left(\hat{c}^{\dagger}\hat{m} + \hat{m}^{\dagger}\hat{c}\right),
    \label{eq:couplingHam}
\end{equation}
where $\omega_c$($\omega_m$) is the resonant frequency of the cavity mode $\hat{c}$ (magnon mode $\hat{m}$) and $G = \sqrt{N}g$ is the coupling strength, proportional to the number of spins $N$ and single ion coupling strength $g$. From the fit we obtain a coupling strength of $G =$ 1.72\,GHz, which divided by the central frequency of the cavity ($f_0 = 11.245$\,GHz) gives a coupling figure of $G/f_0$ = 0.15 putting the system in the ultrastrong coupling regime.

\begin{figure}
  \centering
  \includegraphics[width=1.0\columnwidth]{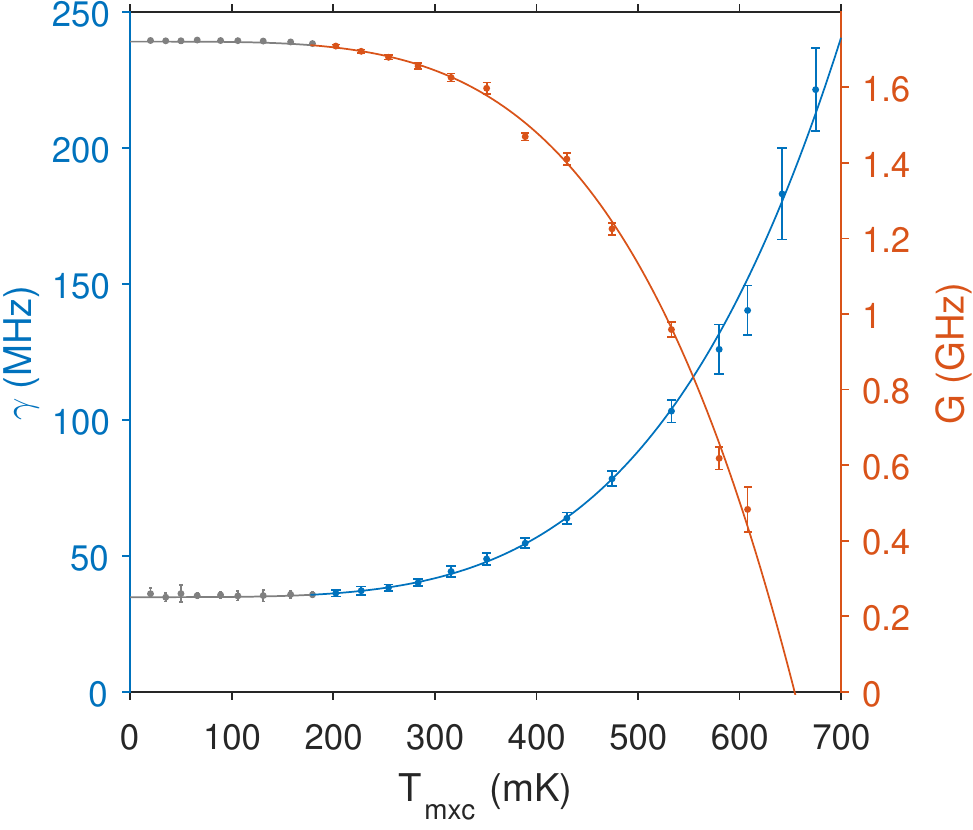}
  \caption{
  \label{fig:gc_and_lw} Blue - polariton linewidth measured at a fixed frequency of 15.6\,GHz as a function of the mixing chamber temperature, Orange - coupling strength $G$ obtained from fitting the coupling Hamiltonian Eq.\eqref{eq:couplingHam}. The grey points indicate the region where we haven't ruled out the possibility that the mixing chamber and the sample are different temperatures.}
\end{figure}

Measuring the linewidth of the polariton modes at points far away from the central cavity frequency gives us an estimate for the linewidth of the antiferromagnetic resonant mode. In order to avoid the background frequency dependence of the experimental components (cables/attenuators/amplifier) we measure the magnetic linewidth of the polariton. The linewidth measured in Tesla is then converted into the frequency domain via the relation $\gamma(\omega) = \gamma(B)g\mu_B$, where $\gamma(\omega)$ and $\gamma(B)$ are the polariton linewidths in the frequency and magnetic domains respectively. Fixing the frequency at 15.6\,GHz the linewidth of the upper branch was measured as a function of the mixing chamber temperature ($T_{\text{mxc}}$), as shown in Fig.~\ref{fig:gc_and_lw}.  As the temperature of the mixing chamber is lowered the magnon linewidth reduces until a temperature of $T_{\text{mxc}} \le 200$\,mK where the linewidth remains roughly constant at $\sim35$\,MHz. The reduction in linewidth with temperature is expected because of a reduction in multimagnon processes \cite{rezende_multimagnon_1976}. The constant linewidth we see for $T_{\text{mxc}} \le 200$\,mK could be due to some non-magnetic broadening mechanism, however it could also be explained by imperfect thermalisation between the mixing chamber and our sample leaving the sample temperature constant at $\sim200$\,mK despite the lower mixing chamber temperature. The smooth line shown behind the linewidth data points is a fit to the equation $\gamma = A + BT^{4}$, where the coefficients were calculated as $A = 34.8$\,MHz and $B = 8.6 \times 10^{-10}$\,MHz/K$^{4}$.

The coupling strength $G$ measured via fitting Eq.~\eqref{eq:couplingHam} is also shown as a function of $T_{\text{mxc}}$ in Fig.~\ref{fig:gc_and_lw}. The coupling strength follows a similar trend to the linewidth data, remaining constant at $\sim1.72$\,GHz up until $T_{\text{mxc}} \approx 200$\,mK, after which it starts decreasing proportional to $T^4$. Fitting the expression $G = A - BT^{4}$ gives coefficients $A = 1.7$\,GHz and $B = 9.39 \times 10^{-12}$\,GHz/K$^{4}$.      

Our results put us in the ultrastrong coupling regime, $G/f_0=0.15>0.1$, the region where counter rotating terms in the coupling Hamiltonian start to become significant. With a lower frequency resonator ($\le 1.7$\,GHz) the deep strong coupling regime could be reached. The challenge will be keeping narrow magnon linewidths in spite of this low frequency which will require low spin temperatures to maintain high spin order. 

Our results have significant implications for microwave to optical conversion, however GdVO$_4$ is unlikely to be the best material for this. The Gd$^{3+}$ ions have half filled $4f$ orbitals and as a result the lowest energy $4f-4f$ transition from the ground-state is in the ultraviolet spectrum around $315$\,nm. Unfortunately the VO$_{4}^{3-}$ ions absorb strongly at this wavelength. Different rare earth crystals with better optical properties for upconversion will be the subject of future work.

Another exciting prospect provided by rare earth ions is the large spectroscopic $g$ factors available, particularly in erbium and dysprosium where $g\approx 15$ is common. With this the collective coupling, which is linear in $g$, could be further improved.

Our measurements show that narrow collective magnetic resonances occur in rare earth crystals.  In comparison to rare earth doped samples the concentration to linewidth ratio seen here is orders of magnitude higher.
Quantum magnonics using rare earth crystals has an exciting future. The remarkable properties of the rare earth $4f-4f$ transitions, the easy accessibility of antiferomagnetic resonances as well as the large $g$ factors, open up many possibilities for improvements to hybrid quantum systems.

\section{Methods}

\begin{figure}[h!]
  \centering
  \includegraphics[width=0.7\columnwidth]{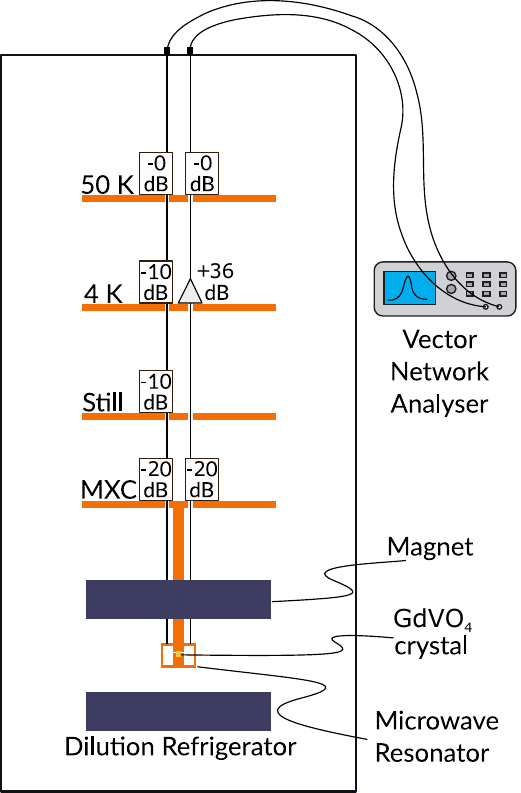}
  \caption{
  \label{fig:setup} Schematic of the experimental setup.}
\end{figure}

The experimental setup used to investigate the coupling between antiferromagnetic modes and a microwave cavity is shown in Fig.~\ref{fig:setup}.  A loop-gap microwave resonator \cite{ball_loop-gap_2018}, with a central frequency of 11.245\,GHz and a Q  factor of 1300, is mounted on the coldest stage of a dilution fridge (BlueFors LD-250) and inside the bore of a 3\,T superconducting magnet. The magnetic field was applied along the crystal $c$-axis, which was identified by observing the crystal through crossed polarisers.

Microwaves are coupled into the resonator using two wire antennas attached to SMA connectors. A vector network analyser is used as the microwave signal source and detector. To reduce the electronic noise the input signal is attenuated at various temperature stages of the dilution fridge, while on the output line a cryoamplifier is used to improve the signal-to-noise ratio. An attenuator is also added between the amplifier and the cavity; this is to suppress heating due to backaction from the input of the amplifier.

\bibliography{gvo_ref}

\section{Acknowledgements}
The authors would like to thank Rose Ahlefeldt, Matt Berrington and Matt Sellars for valuable discussions. This work was supported by Army Research
Office (ARO/LPS) (CQTS) grant number W911NF1810011, the Marsden Fund (Contract No. UOO1520) of the Royal Society of New Zealand, the ARC Centre of Excellence for Quantum Computation and Communication Technology (Grant CE170100012) and the
Discovery Project (Grant DP150103699). 

\end{document}